\documentclass[aps,showpacs,pra,superscriptaddress,]{revtex4-2}
\usepackage{amsmath}
\usepackage{amsfonts}
\usepackage{amssymb}
\usepackage{bm}
\usepackage{epsfig}
\usepackage{graphicx}

\begin{document}
	
\title{Pulse-train propagation in a nonlinear Kerr medium governed by higher-order dispersion}
\author{ Houria Triki}
\affiliation{Radiation Physics Laboratory, Department of Physics, Faculty of Sciences,
Badji Mokhtar University, P. O. Box 12, 23000 Annaba, Algeria}
\author{Vladimir I. Kruglov}
\affiliation{Centre for Engineering Quantum Systems, School of Mathematics and Physics, The University of Queensland, Brisbane, Queensland 4072, Australia}

\title{Pulse-train propagation in a nonlinear Kerr medium governed by
higher-order dispersion}

\begin{abstract}
We discover three novel classes of pulse-trains in an optical Kerr nonlinear
medium possessing all orders of dispersion up to the fourth order. We show
that both single- and double-humped pulse-trains can be formed in the
nonlinear medium. A distinguishing property is that these structures have
different amplitudes, widths and wave numbers but equal velocity which
depends on the three dispersion parameters. More importantly, we find that
the relation between the amplitude and duration of all the newly obtained
pulse-trains is determined by the sign of a joint parameter solely. The
results show that those optical waves are general, in the sense that no
specified conditions on the material parameters are taken on. Considering
the long-wave limit, the derived pulse-trains degenerate to soliton pulses
of the quartic and dipole kinds. Results in this study may be useful for
experimental realization of pulse-trains in highly dispersive optical fibers
and further understanding of their optical transmission properties.

\medskip

Keywords: Pulse-train, nonlinear Kerr medium, Higher-order dispersion,
Quartic and dipole solitons, Nonlinear Schr\"{o}dinger equation.
\end{abstract}

\maketitle

\section{Introduction}

The transmission of envelope solitons in waveguiding media has been the
focus of continuing interest, stimulated by their important potential
applications in optical communication systems and optical switching devises 
\cite{Stegeman}. The formation of such wave packets in a monomode optical
fiber results from an interplay between nonlinear self-phase modulation and
group-velocity dispersion effects \cite{Mihalache}. One notes that the
experimental observation of solitons has been demonstrated in diverse
optical systems, including nonlinear optical fibers \cite{Exp1}, glass
waveguides \cite{Exp3}\ and femtosecond ring lasers \cite{Exp2}. Recently,
the properties of soliton pulses have been studied extensively due to their
important applications for optical transmission and information processing 
\cite{Exp4}.

Usually, the evolution of picosecond light pulses inside a monomode fiber
has been more generally governed by the nonlinear Schr\"{o}dinger equation
(NLSE) that is commonly derived with use of the slowly varying envelope
approximation \cite{A}. Such a pertinent model has admitted the bright pulse
soliton solution in the anomalous dispersion regime while in the normal
dispersion region of the fiber it has exhibited the dark-type soliton
solution \cite{Ha}. However, when much shorter optical pulses have to be
injected in the fiber, many higher-order dispersive and nonlinear effects
have become significant and should have been taken into consideration \cite%
{Zhao}. For example, the third-order dispersion has considerably influenced
the short pulses whose widths are about $50$ femtoseconds as they propagate
through the nonlinear medium \cite{P1,P2}. The quartic dispersion has been
also found to be significant when the light pulses are shorter than $10$
femtoseconds \cite{P1,P2}. In such a case, extensions of the cubic model
taking into account the specific contributions of various higher-order
effects on the light propagation have been therefore essential to describe
the wave dynamics accurately \cite{Kru}. It has been worth to noting that
the occurrence of higher-order processes gives rise to a multitude of novel
phenomena which are not found in the case of the pure Kerr media \cite{Tai}.
One notes that besides the NLSE and its variants, important results have
been also obtained with recent studies discussing diverse nonlinear models
which find applications in various sciences and engineering fields,
including magnetic field \cite{Gao1}, fluid dynamics \cite{Gao2,Gao3},
magneto-optics \cite{Gao8}, quantum mechanics \cite{Gao9}, and plasma
physics \cite{Gao10}. These models have been used to understand and simulate
complex phenomena like solitons in magnetic materials, shallow-water waves
in oceans, and the propagation of nonlinear waves in various media. We also
note that significant results have been recently obtained by studying the
existence and dynamics of various new soliton structures in nonlinear
physical systems governed by the NLS-type equations, thus illustrating their
significance in fields like optics and Bose-Einstein condensates \cite%
{Zhou1,Zhou2,Zhou3}. Techniques like bilinear auto-B\"{a}cklund
transformations and similarity reductions have been found to be crucial for
finding solutions for such models and understanding system behaviors.

Nowadays, a great interest has been given to the generation of pulse-trains
in various physical systems, including birefringent optical fibers \cite%
{Chow1,Chow2,Chow3,H1,H2}, negative-index materials \cite{H3}, Bose-Einstein
condensates \cite{H4}, etc. It can be relevant to note that such type of
pulse shape is attracting in itself as another kind of physically relevant
nonlinear waves in optical fibers besides the soliton waveforms. These
structures exhibit many interesting properties that make them potentially in
problems of signal propagation in nonlinear fibers \cite{Akhmediev}. It is
worth mentioning that pulse-trains can be produced experimentally. For
example, the formation of cnoidal waves has been recently demonstrated in a
fiber laser either in the net anomalous or net normal cavity dispersion
regime \cite{Huu}. Importantly, such pulse-trains may be created from a
dual-frequency pump \cite{Pitois} or by the modulational instability of the
continuous-wave signal \cite{Tomita}. In optics, such nonlinear structures
play a pivotal role in analyzing the data transmission in optical
communication links \cite{Chao}.

Recently, a kind of extended NLSE with second-, third-, and fourth-order
dispersion terms has been shown to support a stable solitary wave solution
with a $\mathrm{sech}^{2}$ shape which is also called quartic soliton \cite%
{Kruglov2}. Such kind of stable localized pulses could find practical
applications in communications, ultrafast lasers, and slow-light devices 
\cite{Kruglov2}. Due to its physical importance in describing femtosecond
pulse propagation in highly dispersive optical fibers, such equation has
been analyzed from different points of view \cite{Kruglov2,Kruglov3,Kruglov4}%
. With regards to the periodic waves, one of the present authors \cite%
{Kruglov4} has analytically solved this model and obtained results for
periodic wave solutions of the $\mathrm{cn}^{2}$ type. However, analytical
pulse-train solutions of this extended NLSE that are expressed in terms of
products of two different Jacobi elliptic functions have not been reported
yet.

Identification of new kinds of pulse-trains in optical systems is an issue
worth considering. It is pertinent to mention that the finding of such
structures may lead to novel dynamical behaviors of waves and opens the
possibility for experimental investigations of various wave phenomena. In
this study, we show for the first time that an optical Kerr nonlinear system
displaying dispersion effects up to the fourth order supports the existence
of three novel classes of pulse-trains which take either single- or
double-hump forms. The main novelty of these solutions is that the periodic
pulse-trains that are composed of products of Jacobian elliptic functions
are firstly reported for the extended NLSE with third- and fourth-order
dispersions, the interesting amplitude-duration relationship for these
structures is found, and the propagation of pulse-train solutions is firstly
represented by numerical simulations. In particular, the outcomes presented
below constitute the first analytical demonstration of possible pulse-train
generation in highly dispersive optical fibers, which may have potential
application for the further experiments and research in nonlinear optics.
These results will be helpful to nonlinear optics since this kind of
nonlinear waves serves as a model of pulse train propagation in optics
fibers \cite{Chao} due to their structural stability with respect to the
small input profile perturbations and collisions \cite{Petnikova}.

The rest of paper is structured as follows. In Sec. II, we present the
analytic framework necessary to find the traveling wave solutions of the
extended NLSE model which governs the transmission of extremely short light
pulses through a fiber system depicting higher-order dispersion effects. In
Sec. III, we derive three different analytic pulse-train solutions of the
governing equation and their characteristics. We also discuss here the
propagation properties of obtained pulse-trains in the fiber medium. In Sec.
IV, we numerically simulate the propagation of the obtained pulse-train
solutions. The stability of the solutions is discussed numerically in Sec.
V. The conclusions of this paper will be then summarized in Sec. VI.

\section{Model and traveling waves}

We consider the propagation of ultrashort light pulses through a nonlinear
fiber medium exhibiting the effects of higher-order dispersions as well as
self-phase modulation nonlinearity. The governing model for the slowly
varying envelope $\psi $ of the light pulse can be described by the extended
NLSE \cite{Kruglov2,Kruglov3}, 
\begin{equation}
\mathrm{i}\frac{\partial \psi }{\partial z}=\alpha \frac{\partial ^{2}\psi }{%
\partial \tau ^{2}}+\mathrm{i}\text{\/}\sigma \frac{\partial ^{3}\psi }{%
\partial \tau ^{3}}-\epsilon \frac{\partial ^{4}\psi }{\partial \tau ^{4}}%
-\gamma \left\vert \psi \right\vert ^{2}\psi ,  \label{1}
\end{equation}%
where $z$ is the longitudinal coordinate, $\tau =t-\beta _{1}z$ is the
retarded time, $\alpha =\beta _{2}/2$, $\sigma =\beta _{3}/6$, $\epsilon
=\beta _{4}/24$, while $\gamma $ is the fiber nonlinearity coefficient that
results in self-phase modulation. Here $\beta _{j}$ ($j=1,2,3,4$) represent,
respectively, the dispersion coefficients of the first, second, third, and
fourth order.

As an important generalization, the extended NLSE (\ref{1}) includes,
besides the group-velocity dispersion and self-phase modulation terms that
constitute the cubic NLS model, the effects of third- and fourth-order
dispersions. These higher-order effects arise in highly dispersive optical
systems including dispersion-shifted fibers \cite{Boggio}, photonic crystal
waveguides \cite{Assefa}, and silicon-based waveguides \cite{Roy2}. It is
relevant to note here that the third- and fourth-order dispersions occur in
the region of minimum group-velocity dispersion \cite{Cavalcanti}, in which
they become crucial for describing femtosecond pulse behavior.

Theoretical investigations based on Eq. (\ref{1}) are mainly concentrated on
the existence and propagation properties of localized pulses due to their
fundamental importance in the understanding of various physical phenomena in
the system. As previously mentioned, Kruglov and Harvey \cite{Kruglov2}
demonstrated the existence of stable solitary waves with a $\mathrm{sech}%
^{2} $ shape for this model. Triki and Kruglov \cite{Kruglov3} discussed the
dynamics of dipole soliton waveforms of Eq. (\ref{1}) in the presence the
inhomogeneities of media. Kruglov \cite{Kruglov4} obtained the periodic wave
solutions that take the $\mathrm{cn}^{2}$ shape for this underlying equation 
\cite{Kruglov4}. Cavalcanti et al. \cite{Cavalcanti} analyzed the modulation
instability of the model (\ref{1}) in the region near the zero-dispersion
wavelength. Karpman et al. \cite{K1,K2} investigated the resonant radiation
and evolution of a soliton described by the model (\ref{1}). Shagalov \cite%
{Shagalov} studied the influence of high dispersion terms in the governing
equation (\ref{1})\ on the modulational instability of nonlinear waves. Roy
et al. \cite{Roy} discussed the roles of high-order dispersions in the
generation and control of dispersive waves. In what follows,\ we present
three novel analytic pulse-train solutions with interesting properties for
the extended NLSE (\ref{1}), which are obtained without necessarily assuming
a specified condition on the fiber parameters. Here, the ansatz method which
is one of the effective and powerful techniques for finding analytic
localized and periodic wave solutions of nonlinear evolution equations \cite%
{Gao1,Gao2,Gao3}, is used to obtain various types of pulse-trains solutions
for the studied model.

In order to find some interesting analytical solutions of the model (\ref{1}%
), we make the transformation \cite{Kruglov2,Kruglov3}, 
\begin{equation}
\psi (z,\tau )=\rho (\xi )\exp [\mathrm{i}(\kappa z-\delta \tau +\theta )],
\label{2}
\end{equation}%
with $\rho (\xi )$ is the real amplitude and $\xi =\tau -qz$ is the
traveling coordinate. Here the inverse velocity $q=v^{-1}$, the frequency
shift $\delta $, the wave number $\kappa $, and the phase constant $\theta $
(at $z=0$) are all real parameters to be determined.

Inserting Eq. (\ref{2}) into the extended NLSE (\ref{1}), we get the
differential equations:%
\begin{equation}
(\sigma +4\epsilon \delta )\frac{\mathrm{d}^{3}\rho }{\mathrm{d}\xi ^{3}}%
+(q-2\alpha \delta -3\sigma \delta ^{2}-4\epsilon \delta ^{3})\frac{\mathrm{d%
}\rho }{\mathrm{d}\xi }=0,  \label{3}
\end{equation}%
\begin{equation}
\epsilon \frac{\mathrm{d}^{4}\rho }{\mathrm{d}\xi ^{4}}-(\alpha +3\sigma
\delta +6\epsilon \delta ^{2})\frac{\mathrm{d}^{2}\rho }{\mathrm{d}\xi ^{2}}%
+\gamma \rho ^{3}-(\kappa -\alpha \delta ^{2}-\sigma \delta ^{3}-\epsilon
\delta ^{4})\rho =0.  \label{4}
\end{equation}%
Here, one refers the waveform solution of the model under consideration
where the envelope function $\psi (z,\tau )$ is defined by the
representation (\ref{2}) with $\rho (\xi )\neq \mathrm{constant}$ as
traveling wave or non-plain wave solution.

Equations (\ref{3}) and (\ref{4}) with non-zero quartic dispersion
coefficient (i.e., $\epsilon \neq 0$) give the non-plain wave solutions when
the following relations are fulfilled: 
\begin{equation}
q=2\alpha \delta +3\sigma \delta ^{2}+4\epsilon \delta ^{3},~~~~\delta =-%
\frac{\sigma }{4\epsilon }.  \label{5}
\end{equation}%
Notice that the differential equation (\ref{3}) is fulfilled for an
arbitrary amplitude function $\rho (\xi )$ in accordance with the conditions
in Eq. (\ref{5}) with $\epsilon \neq 0$. Additionally, Eq. (\ref{5}) allows
us to obtain the following expression for the wave velocity $v=1/q$,%
\begin{equation}
v=\frac{8\epsilon ^{2}}{\sigma (\sigma ^{2}-4\alpha \epsilon )}.  \label{6}
\end{equation}

The latter relation shows that the dispersion parameters $\beta _{i}$ $%
(i=2,3,4)$ included in the parameters $\alpha $, $\sigma $ and $\epsilon $
have a major influence on the wave velocity. This implies that the velocity
of the pulse can be efficiently controlled by varying these parameters.

Now using Eqs. (\ref{4}) and (\ref{5}), one can find an equation for $\rho $
as 
\begin{equation}
\epsilon \frac{\mathrm{d}^{4}\rho }{\mathrm{d}\xi ^{4}}+b\frac{\mathrm{d}%
^{2}\rho }{\mathrm{d}\xi ^{2}}-c\rho +\gamma \rho ^{3}=0,~~~~~~~  \label{7}
\end{equation}%
with 
\begin{equation}
b=\frac{3\sigma ^{2}}{8\epsilon }-\alpha ,~~~~c=\kappa +\frac{\sigma ^{2}}{%
16\epsilon ^{2}}\left( \frac{3\sigma ^{2}}{16\epsilon }-\alpha \right) .
\label{8}
\end{equation}

Equation (\ref{7}) describes the evolution of the wave amplitude in the
highly dispersive nonlinear fiber. Now the natural question is whether the
amplitude equation (\ref{7}) possesses analytic pulse-train solutions
without any specified condition on the coefficients or not. In following, we
have identified novel families of pulse-train solutions with some
interesting properties for this evolution equation taking into account the
influence of all linear and nonlinear processes on pulse propagation.

\section{Results and discussion}

In this section, we demonstrate that new kinds of pulse-trains can exist in
the fiber system governed by Eq. (\ref{1}). Remarkably, we find that these
propagating waves possess distinct amplitudes and consequently different
powers, which is highly desired from a practical point of view.\newline

\subsection{Pulse-train solutions}

Here we show for the first time that Eq. (\ref{7}) allows for analytic
pulse-train solutions which are expressed in terms of the product of
different elliptic functions. We find that this type of pulse shape can
exist for Eq. (\ref{1}) in three different classes.

\begin{figure}[h]
\includegraphics[width=1\textwidth]{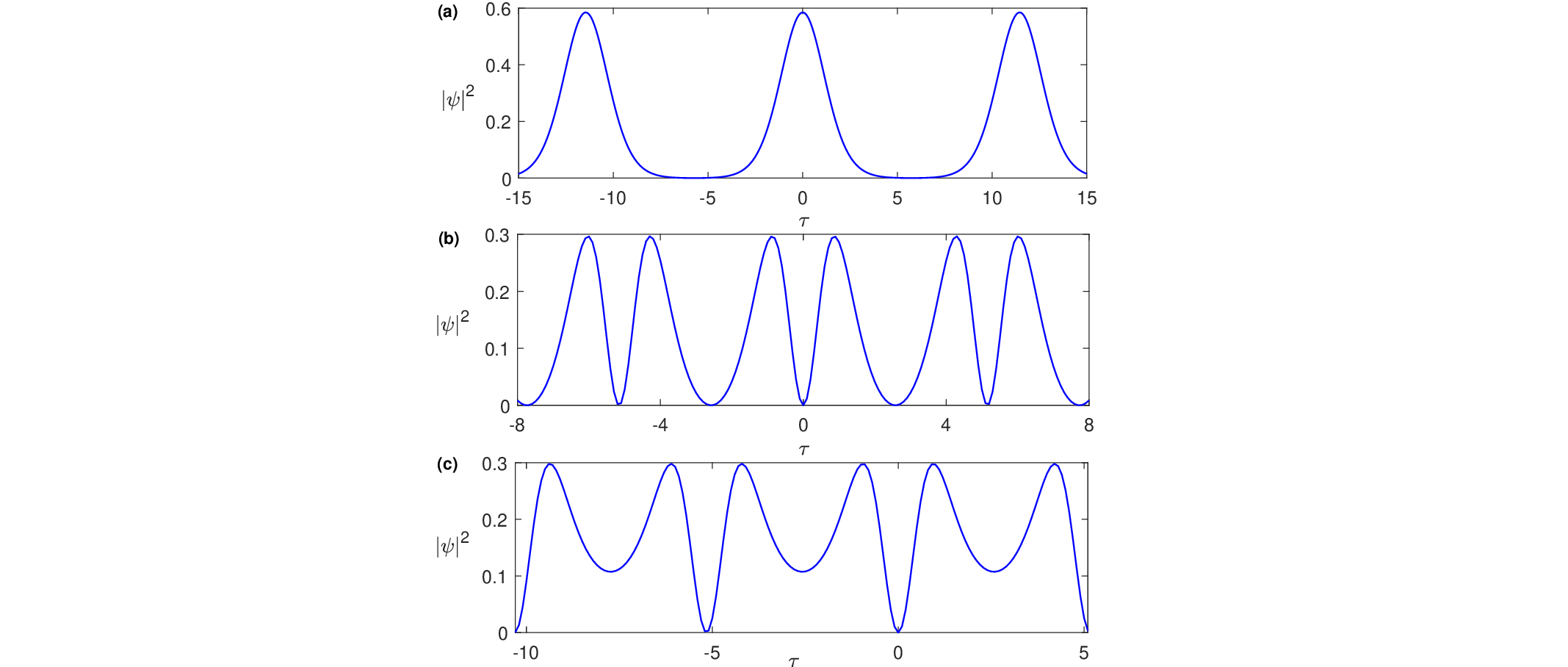}
\caption{Intensity profiles of pulse-train solutions with parameters $%
\protect\gamma =2$, $k=0.9$, $\protect\xi _{0}=0$; (a) pulse-train solution (%
\protect\ref{16}) when $\protect\alpha =-1,$ $\protect\sigma =-0.25,$ $%
\protect\epsilon =-0.25$; (b) pulse-train solution (\protect\ref{25}) when $%
\protect\alpha =0.49,$\ $\protect\sigma =1,$ $\protect\epsilon =0.5$; (c)
pulse-train solution (\protect\ref{34}) when $\protect\alpha =0.49,$\ $%
\protect\sigma =1,$ $\protect\epsilon =0.5.$ }
\label{FIG.1.}
\end{figure}

\begin{figure}[h]
\includegraphics[width=1.3\textwidth]{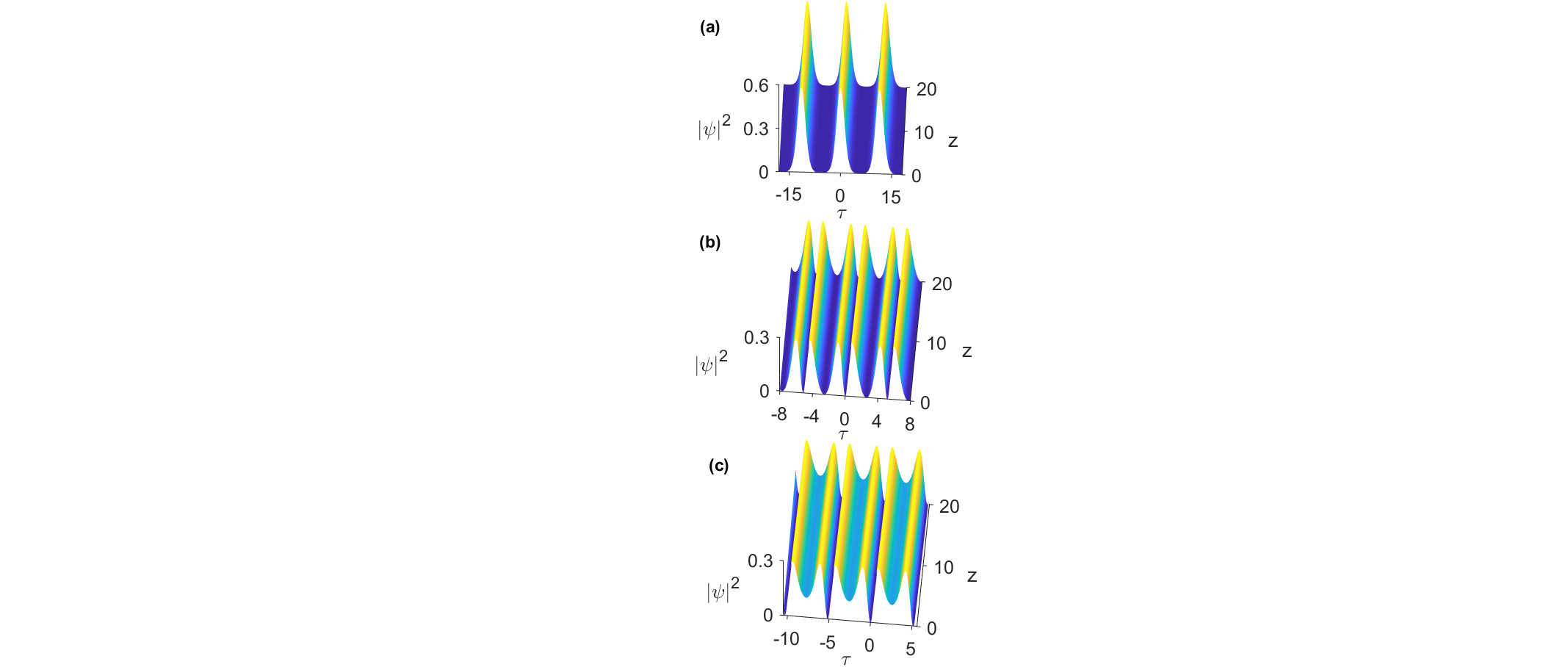}
\caption{Intensity profiles of pulse-train solutions with parameter $k=0.9$;
(a) pulse-train solution (\protect\ref{16}), (b) pulse-train solution (%
\protect\ref{25}), (c) pulse-train solution (\protect\ref{34}). Other
parameters are the same as given in Fig. 1. }
\label{FIG.2.}
\end{figure}

\begin{figure}[h]
\includegraphics[width=1\textwidth]{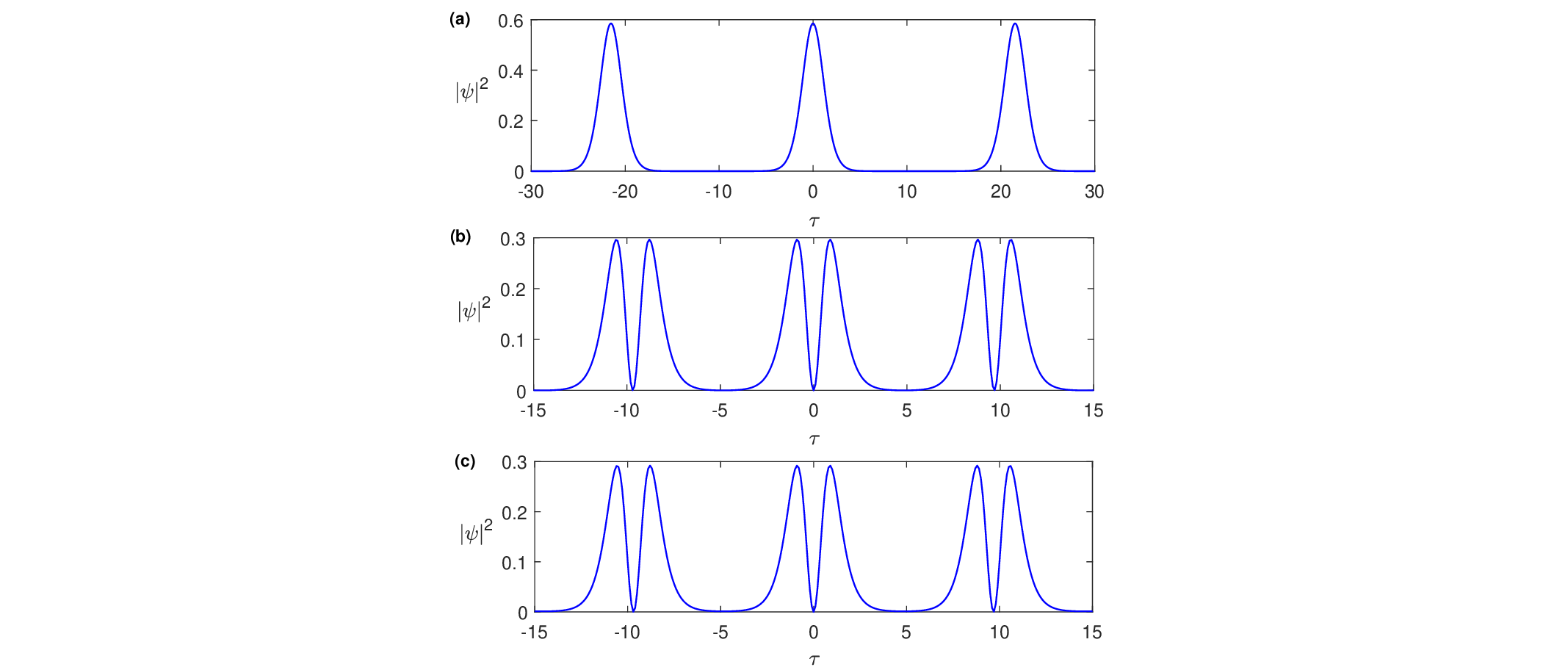}
\caption{Evolution of pulse-train solutions with parameter $k=0.999$; (a)
pulse-train solution (\protect\ref{16}), (b) pulse-train solution (\protect
\ref{25}), (c) pulse-train solution (\protect\ref{34}). Other parameters are
the same as given in Fig. 1. }
\label{FIG.3.}
\end{figure}
\begin{figure}[h]
\includegraphics[width=1.3\textwidth]{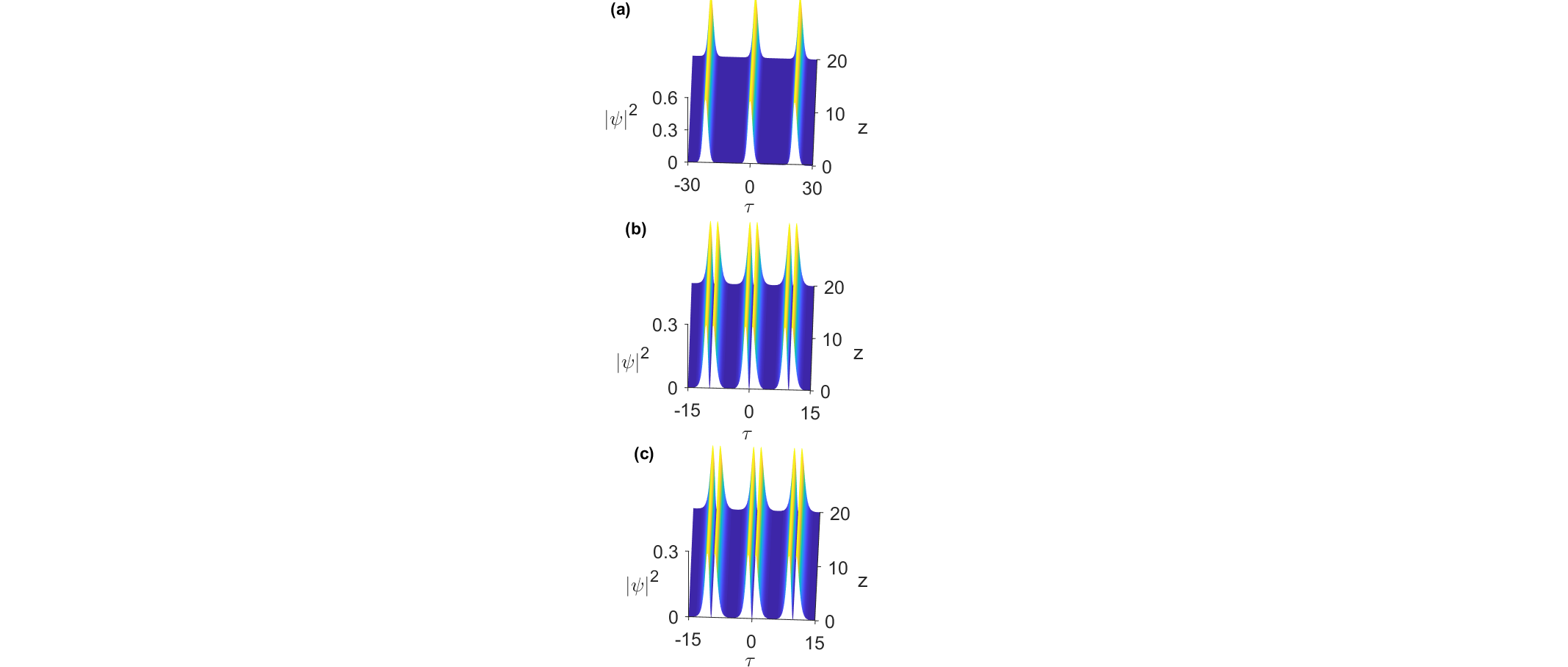}
\caption{Evolution of pulse-train solutions with parameter $k=0.999$; (a)
pulse-train solution (\protect\ref{16}), (b) pulse-train solution (\protect
\ref{25}), (c) pulse-train solution (\protect\ref{34}). Other parameters are
the same as given in Fig. 1.}
\label{FIG.4.}
\end{figure}

\begin{figure}[h]
\includegraphics[width=1\textwidth]{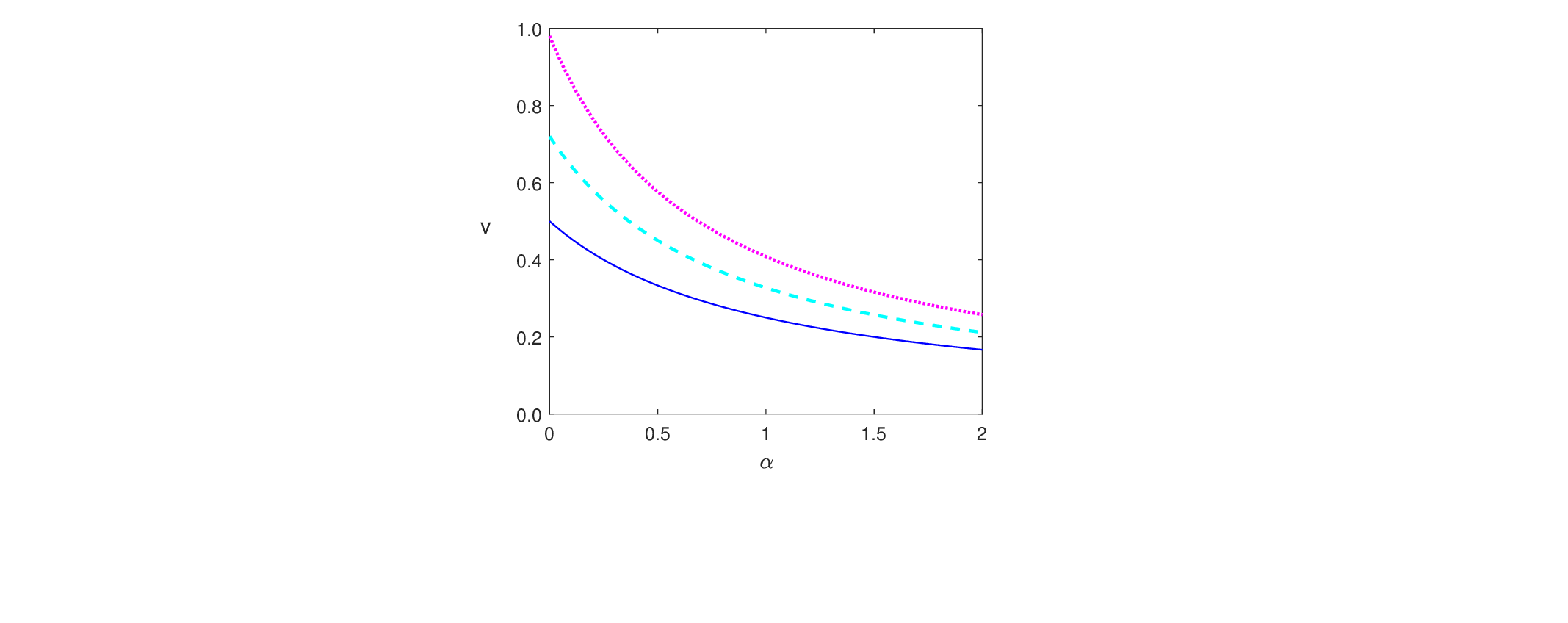}
\caption{Dependence of the velocity of pulse-trains $v$ on the group
velocity dispersion parameter $\protect\alpha $ for $\protect\sigma =1$ and
different values of $\protect\epsilon $; $\protect\epsilon =-0.25$ (thick
line), $\protect\epsilon =-0.3$ (dashed line) and $\protect\epsilon =-0.35$
(dotted line). }
\label{FIG.5.}
\end{figure}

\begin{figure}[h]
\includegraphics[width=1.3\textwidth]{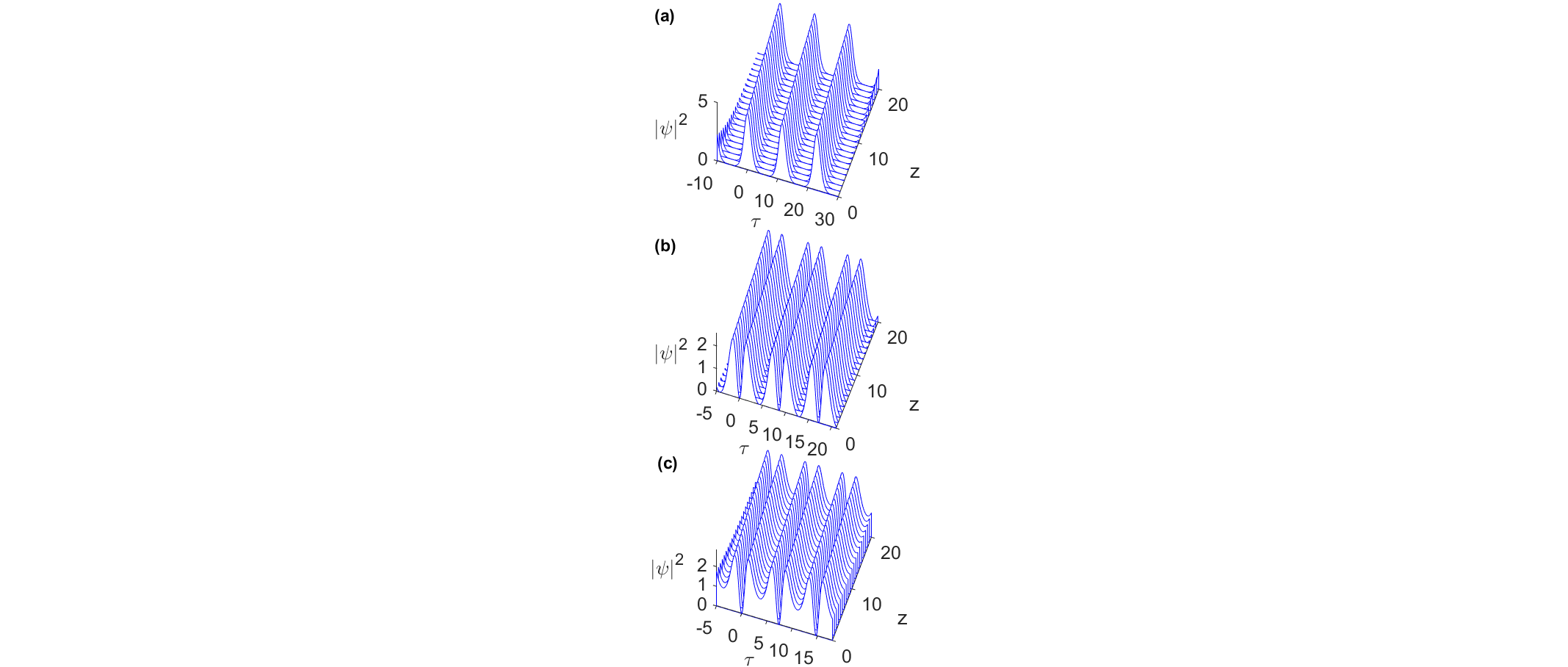}
\caption{Evolution of the intensity distribution of the extended NLSE; (a)
pulse-train solution (\protect\ref{16}), (b) pulse-train solution (\protect
\ref{25}), (c) pulse-train solution (\protect\ref{34}). Parameters are given
in the text.}
\label{FIG.6.}
\end{figure}

\begin{description}
\item[\textbf{1. Pulse-trains of the }$\mathrm{cn}\left( x,k\right) \mathrm{%
dn}\left( x,k\right) $\textit{\textrm{-}}\textbf{type}] 
\end{description}

We introduce an appropriate ansatz for the propagating waveform solution of
Eq. (\ref{7}) as 
\begin{equation}
\rho (\xi )=P\,\mathrm{dn}\left( w(\xi -\xi _{0}),k\right) \mathrm{cn}\left(
w(\xi -\xi _{0}),k\right) ,  \label{9}
\end{equation}%
with cosine and the third kind of Jacobian elliptic functions $\mathrm{cn}%
\left( w(\xi -\xi _{0}),k\right) $ and $\mathrm{dn}\left( w(\xi -\xi
_{0}),k\right) $ of the modulus number $k$\ ($0<k<1$). Here $P$ is the
amplitude of the wave to be determined and $\xi _{0}$ is its initial
position (at $z=0$).

Substituting the ansatz (\ref{9}) into Eq. (\ref{7}) and setting the
coefficients of $\mathrm{dn}^{j}\left( w(\xi -\xi _{0}),k\right) \mathrm{cn}%
\left( w(\xi -\xi _{0}),k\right) $ functions (where $j=1,3,5$) equal to
zero, one gets the algebraic equations:%
\begin{equation}
\epsilon w^{4}(k^{4}-46k^{2}+61)-bw^{2}(k^{2}-5)=c,  \label{10}
\end{equation}%
\begin{equation}
60\epsilon w^{4}k^{2}(k^{2}-3)-6bw^{2}k^{2}+\gamma (k^{2}-1)P^{2}=0,
\label{11}
\end{equation}%
\begin{equation}
120\epsilon w^{4}k^{2}+\gamma P^{2}=0.  \label{12}
\end{equation}%
By solving the parametric equations (\ref{11}) and (\ref{12}), we get 
\begin{equation}
w=\frac{1}{4}\sqrt{\frac{8\alpha \epsilon -3\sigma ^{2}}{5\epsilon
^{2}(k^{2}+1)}},~~~~P=\pm \frac{k}{k^{2}+1}\sqrt{\frac{-6}{5\gamma \epsilon }%
}\left( \frac{3\sigma ^{2}}{8\epsilon }-\alpha \right),  \label{13}
\end{equation}%
with $\gamma \epsilon <0$ and $8\alpha \epsilon >3\sigma ^{2}$. Further
substitution of Eqs. (\ref{13}) into Eq. (\ref{10}) yields the parameter $c$
as 
\begin{equation}
c=\frac{(11k^{4}-86k^{2}+11)}{100\epsilon (k^{2}+1)^{2}}\left( \frac{3\sigma
^{2}}{8\epsilon }-\alpha \right) ^{2}.  \label{14}
\end{equation}%
Then using Eqs. (\ref{8}) and (\ref{14}), we can obtain the expression for
the wave number $\kappa $: 
\begin{equation}
\kappa =\frac{(11k^{4}-86k^{2}+11)}{100\epsilon ^{3}(k^{2}+1)^{2}}\left( 
\frac{3\sigma ^{2}}{8}-\alpha \epsilon \right) ^{2}-\frac{\sigma ^{2}}{%
16\epsilon ^{3}}\left( \frac{3\sigma ^{2}}{16}-\alpha \epsilon \right) .
\label{15}
\end{equation}

Based on these findings, we get a novel periodic wave solution for the model
(\ref{1}) as 
\begin{equation}
\psi (z,\tau )=P\,\mathrm{dn}\left( w\zeta ,k\right) \mathrm{cn}\left(
w\zeta ,k\right) \exp [\mathrm{i}(\kappa z-\delta \tau +\theta )],
\label{16}
\end{equation}%
where $\zeta =\tau -v^{-1}z-\xi _{0}$. Note that the frequency shift $\delta 
$ and velocity $v$ of this periodic waveform are defined by the relations (%
\ref{5}) and (\ref{6}), respectively.

From Eq. (\ref{13}), one can see that the amplitude $P$ and inverse temporal
width $w$\ of this periodic wave crucially depend on various dispersion
effects as well as nonlinearity. This result indicates that the
contributions of dispersions and self-phase modulation processes are
essential for this newly found pulse-train to be formed in the optical fiber
medium.

We now introduce the quantity $\tau _{0}=1/w$ which represents the duration
of the pulse \cite{Sh}. Hence, we find that the relation between the pulse
amplitude and duration is given by%
\begin{equation}
P\tau _{0}^{2}=-2k\left( -\Omega \right) ^{1/2},  \label{17}
\end{equation}

\noindent where the parameter $\Omega =30\epsilon /\gamma $ has been
introduced for brevity. Note that here, we considered the $+$ sign in all
the expressions of the wave amplitude.

\begin{description}
\item[\textbf{2. Pulse-trains of the }$\mathrm{cn}\left( x,k\right) \mathrm{%
sn}\left( x,k\right) $\textit{\textrm{-}}\textbf{type}] 
\end{description}

We now make an ansatz for Eq. (\ref{7}) as 
\begin{equation}
\rho (\xi )=P\,\mathrm{cn}\left( w(\xi -\xi _{0}),k\right) \mathrm{sn}\left(
w(\xi -\xi _{0}),k\right) ,  \label{18}
\end{equation}%
with cosine and sine Jacobian elliptic functions $\mathrm{cn}\left( w(\xi
-\xi _{0}),k\right) $ and $\mathrm{sn}\left( w(\xi -\xi _{0}),k\right) $ of
the modulus number $k$\ ($0<k<1$).

Upon substituting of this ansatz into Eq. (\ref{7}) and then setting the
coefficients of $\,\mathrm{cn}^{j}\left( w(\xi -\xi _{0}),k\right) \mathrm{sn%
}\left( w(\xi -\xi _{0}),k\right) $ functions (where $j=1,3,5$) equal to
zero, we obtain the algebraic equations: 
\begin{equation}
\epsilon w^{4}(61k^{4}-76k^{2}+16)+bw^{2}(5k^{2}-4)=c,  \label{19}
\end{equation}%
\begin{equation}
60\epsilon w^{4}k^{2}(3k^{2}-2)+6bw^{2}k^{2}+\gamma P^{2}=0,  \label{20}
\end{equation}%
\begin{equation}
120\epsilon \eta ^{4}k^{2}-\gamma P^{2}=0.  \label{21}
\end{equation}%
Based on these equations, one obtains the relations for the inverse width $w$
of the wave and its amplitude $P$ as 
\begin{equation}
w=\frac{1}{4}\sqrt{\frac{3\sigma ^{2}-8\alpha \epsilon }{5\epsilon
^{2}(4-3k^{2})}},~~~~P=\pm \frac{k}{4-3k^{2}}\sqrt{\frac{6}{5\gamma \epsilon 
}}\left( \frac{3\sigma ^{2}}{8\epsilon }-\alpha \right),  \label{22}
\end{equation}%
with $\gamma \epsilon >0$ and $3\sigma ^{2}>8\alpha \epsilon $.
Additionally, by using Eqs. (\ref{19}) and (\ref{22}), we have the
expression:%
\begin{equation}
c=\frac{b^{2}(244k^{2}-89k^{4}-144)}{100\epsilon (4-3k^{2})^{2}}.  \label{23}
\end{equation}

Then from Eqs. (\ref{8}) and (\ref{23}), we find that the wave number $%
\kappa $ of the periodic wave solution takes the form as 
\begin{equation}
\kappa =\frac{(244k^{2}-89k^{4}-144)}{100\epsilon ^{3}(4-3k^{2})^{2}}\left( 
\frac{3\sigma ^{2}}{8}-\alpha \epsilon \right) ^{2}-\frac{\sigma ^{2}}{%
16\epsilon ^{3}}\left( \frac{3\sigma ^{2}}{16}-\alpha \epsilon \right) .
\label{24}
\end{equation}%
These results yield a new kind of pulse-train solution for the NLSE model (%
\ref{1}) as%
\begin{equation}
\psi (z,\tau )=P\,\mathrm{cn}\left( w\zeta ,k\right) \mathrm{sn}\left(
w\zeta ,k\right) \exp [\mathrm{i}(\kappa z-\delta \tau +\theta )],
\label{25}
\end{equation}%
where $\xi =\tau -v^{-1}z-\xi _{0}$. Noting that the frequency shift $\delta 
$ and velocity $v$ in this periodic structure are defined by the relations (%
\ref{5}) and (\ref{6}), respectively.

For this solution, we find that the relation between the pulse amplitude and
duration takes the form,

\begin{equation}
P\tau _{0}^{2}=2k\left( \Omega \right) ^{1/2}.  \label{26}
\end{equation}

\begin{description}
\item[\textbf{3. Pulse-trains of the }$\mathrm{dn}\left( x,k\right) \mathrm{%
sn}\left( x,k\right) $\textit{\textrm{-}}\textbf{type}] 
\end{description}

Next we make an ansatz for Eq. (\ref{7}) as 
\begin{equation}
\rho (\xi )=P\,\mathrm{dn}\left( w(\xi -\xi _{0}),k\right) \mathrm{sn}\left(
w(\xi -\xi _{0}),k\right).  \label{27}
\end{equation}

Substituting this ansatz into the amplitude equation (\ref{7}) and then
setting the coefficients of $\,\mathrm{dn}^{j}\left( w(\xi -\xi
_{0}),k\right) \mathrm{sn}\left( w(\xi -\xi _{0}),k\right) $ functions
(where $j=1,3,5$) equal to zero, one finds the algebraic equations:%
\begin{equation}
\epsilon w^{4}(16k^{4}-76k^{2}+61)+bw^{2}(5k^{2}-4)=c,  \label{28}
\end{equation}%
\begin{equation}
60\epsilon w^{4}k^{2}(2k^{2}-3)-6bw^{2}k^{2}+\gamma P^{2}=0,  \label{29}
\end{equation}%
\begin{equation}
120\epsilon w^{4}k^{2}-\gamma P^{2}=0.  \label{30}
\end{equation}%
Solving these equations, one can get the pulse parameters $w$ and $P$ as 
\begin{equation}
w=\frac{1}{4}\sqrt{\frac{3\sigma ^{2}-8\alpha \epsilon }{5\epsilon
^{2}(2k^{2}-1)}},~~~~P=\pm \frac{k}{2k^{2}-1}\sqrt{\frac{6}{5\gamma \epsilon 
}}\left( \frac{3\sigma ^{2}}{8\epsilon }-\alpha \right) ,  \label{31}
\end{equation}%
with $\gamma \epsilon >0$ and $8\alpha \epsilon <3\sigma ^{2}$ when $1/\sqrt{%
2}<k<1$ and $8\alpha \epsilon >3\sigma ^{2}$ when $0<k<1/\sqrt{2}$.

Moreover, the use of Eqs. (\ref{28}) and (\ref{31}) lead to an expression
for the parameter $c\ $as%
\begin{equation}
c=\frac{b^{2}(64k^{2}-64k^{4}+11)}{100\epsilon (2k^{2}-1)^{2}}.  \label{32}
\end{equation}

Now using Eqs. (\ref{8}) and (\ref{32}), we can determine the wave number $%
\kappa $ as%
\begin{equation}
\kappa =\frac{(64k^{2}-64k^{4}+11)}{100\epsilon ^{3}(2k^{2}-1)^{2}}\left( 
\frac{3\sigma ^{2}}{8}-\alpha \epsilon \right) ^{2}-\frac{\sigma ^{2}}{%
16\epsilon ^{3}}\left( \frac{3\sigma ^{2}}{16}-\alpha \epsilon \right) .
\label{33}
\end{equation}%
Thus, we find a novel pulse-train solution for the NLSE model (\ref{1}) as%
\begin{equation}
\psi (z,\tau )=P\,\mathrm{dn}\left( w\zeta ,k\right) \mathrm{sn}\left(
w\zeta ,k\right) \exp [\mathrm{i}(\kappa z-\delta \tau +\theta )],
\label{34}
\end{equation}%
where $\xi =\tau -v^{-1}z-\xi _{0}$. Here $\delta $ and $v$ can be worked
out by using the relations (\ref{5}) and (\ref{6}), respectively.

For this solution, we find that the relation between the pulse amplitude and
duration takes the form,%
\begin{equation}
P\tau _{0}^{2}=2k\left( \Omega \right) ^{1/2}.  \label{35}
\end{equation}

On comparing Eqs. (\ref{17}), (\ref{26}) and (\ref{35}), we infer that under
the effects of quadratic, cubic and quartic dispersions and self-phase
modulation nonlinearity, the fiber system allows the propagation of distinct
types of pulse-trains which present a relation between the pulse amplitude
and duration determined by the sign of the joint parameter $\Omega
=30\epsilon /\gamma $ only. To our knowledge, the pulse-trains derived above
for Eq. (\ref{1}) are firstly presented in this work.

Figure 1(a) depicts the intensity profile of the pulse-train solution (\ref%
{16}) of the model (\ref{1}) for the parameters values: $\alpha =-1,$\ $%
\gamma =2,$ $\sigma =-0.25,$ and $\epsilon =-0.25.$ The results for the
pulse-train solutions (\ref{25}) and (\ref{34}) are shown in Figs. 1(b) and
1(c) for the values $\alpha =0.49,$\ $\gamma =2,$ $\sigma =1,$ and $\epsilon
=0.5$. One notes here that the value of initial position $\xi _{0}$ (at $z=0$%
) of the nonlinear waves is chosen zero. Also, the modulus number $k$ is
selected to be $k=0.9$.\ The evolution of these pulse-train solutions in the
fiber medium are displayed in Fig. 2. From these figures, we observe that
the three types of pulse-train solutions are positive and exhibit an
oscillating behavior. In addition, we can see that the pulse-train solutions
(\ref{25}) and (\ref{34}) \ present two peaks in each periodic unit,
markedly different from the pulse train solution (\ref{16}) which has a one
peak structure.

It is also interesting to see from Figs. (3) and (4) that the shape of the
pulse-train solution (\ref{25}) is similar to that for the pulse-train
solution (\ref{34}) for the modulus number $k=0.999$.

A key observation is that the amplitudes, widths and wave numbers of the
derived pulse-train solutions are different whereas their velocity is the
same. More importantly, the expression (\ref{6}) demonstrates that the
velocity of these pulse-trains can be considerably diminished for suitable
parameters of the waveguiding media, which have application value in the
development of slow-light systems. Recently, Li and Huang \cite{Huang}
studied the modification of a slow-light soliton in three-level atomic
systems. They derived a high-order NLSE incorporating third-order
dispersion, delay in nonlinear refractive index, modulation induced by the
microwave field, nonlinear dispersion, and correction terms of linear and
differential absorptions; and demonstrated that an obvious decrease of
propagating velocity of the soliton can be obtained in the presence of the
microwave field. Note that photonic crystal waveguides (PhCWs) represent
interesting slow light structures due to their confinement and versatile
dispersion properties including slow light and possible control of group
velocity dispersion effects \cite{Ph1,Ph2,Ph3}. In order to strictly answer
the question of possibility of applying the newly found pulse-trains in the
development of slow-light systems much further research is needed. A
detailed investigation of this problem will be reported in a forthcoming
publication.

In Fig. 5, we present the dependence of velocity on the group velocity
dispersion parameter $\alpha $ of the medium for three distinct values of
quartic dispersion coefficient $\epsilon $. We clearly see that the wave
velocity decreases intensely in the magnitude as the value of dispersion
parameter $\alpha $ is increased. The decrease of velocity implies that the
slowing motion of pulse-trains is possible in the transmitting medium when
considering large dispersion parameter $\alpha $. It is also observed that
the pulse velocity has large magnitude for small negative values of\ $%
\epsilon $. Hence, one concludes that the velocity of pulse-trains can be
controlled by judicious choice of the dispersion coefficients.

\subsection{Quartic and dipole soliton solutions}

We now focus on the long-wave limit of the new pulse-trains obtained above
corresponding to the modulus $k=1$. In such a case, the periodic solution (%
\ref{16}) degenerates to the quartic soliton solution \cite{Kruglov2}:%
\begin{equation}
\psi (z,\tau )=P_{0}~\mathrm{sech}^{2}\left[ w_{0}\left( \tau -v^{-1}z-\xi
_{0}\right) \right] \exp [\mathrm{i}(\kappa z-\delta \tau +\theta )],
\label{36}
\end{equation}%
where the inverse temporal width $w_{0}$, amplitude $P_{0}$ and wave number $%
\kappa $ of the soliton pulse are given by%
\begin{equation}
w_{0}=\frac{1}{4}\sqrt{\frac{8\alpha \epsilon -3\sigma ^{2}}{10\epsilon ^{2}}%
},  \label{37}
\end{equation}%
\begin{equation}
P_{0}=\pm \sqrt{\frac{-3}{10\gamma \epsilon }}\left( \frac{3\sigma ^{2}}{%
8\epsilon }-\alpha \right) ,  \label{38}
\end{equation}%
\begin{equation}
\kappa =-\frac{4}{25\epsilon ^{3}}\left( \frac{3\sigma ^{2}}{8}-\alpha
\epsilon \right) ^{2}-\frac{\sigma ^{2}}{16\epsilon ^{3}}\left( \frac{%
3\sigma ^{2}}{16}-\alpha \epsilon \right) ,  \label{39}
\end{equation}

\noindent with $\gamma \epsilon <0$ and $8\alpha \epsilon >3\sigma ^{2}$.
Here the soliton frequency shift $\delta $ is given by Eq. (\ref{5}) and the
velocity $v$ is defined by Eq. (\ref{6}).

Notice that for this quartic soliton waveform, the relation between the
amplitude and duration is given by%
\begin{equation}
P_{0}\tau _{0}^{2}=-2\sqrt{\frac{-30\epsilon }{\gamma }}.  \label{40}
\end{equation}

\noindent provided that $\gamma \epsilon <0$.

In addition, when $k\rightarrow 1$, the periodic solution (\ref{25})
transforms into our dipole soliton solution \cite{Kruglov3}:%
\begin{equation}
\psi (z,\tau )=P_{0}~\mathrm{sech}\left[ w_{0}\left( \tau -v^{-1}z-\xi
_{0}\right) \right] \mathrm{th}\left[ w_{0}\left( \tau -v^{-1}z-\xi
_{0}\right) \right] \exp [\mathrm{i}(\kappa z-\delta \tau +\theta )],
\label{41}
\end{equation}%
where the inverse temporal width $w_{0}$, amplitude $P_{0}$ and wave number $%
\kappa $ of the soliton pulse take the form, 
\begin{equation}
w_{0}=\frac{1}{4}\sqrt{\frac{3\sigma ^{2}-8\alpha \epsilon }{5\epsilon ^{2}}}%
,  \label{42}
\end{equation}%
\begin{equation}
P_{0}=\pm \sqrt{\frac{6}{5\gamma \epsilon }}\left( \frac{3\sigma ^{2}}{%
8\epsilon }-\alpha \right) ,  \label{43}
\end{equation}%
\begin{equation}
\kappa =\frac{11}{100\epsilon ^{3}}\left( \frac{3\sigma ^{2}}{8}-\alpha
\epsilon \right) ^{2}-\frac{\sigma ^{2}}{16\epsilon ^{3}}\left( \frac{%
3\sigma ^{2}}{16}-\alpha \epsilon \right) ,  \label{44}
\end{equation}%
with $\gamma \epsilon >0$ and $8\alpha \epsilon <3\sigma ^{2}$. Notice that
the frequency shift $\delta $ of this dipole soliton is given by Eq. (\ref{5}%
) and its velocity $v$ is defined by Eq. (\ref{6}).

It is interesting to note that if one considers the limiting case with $k=1$%
, the pulse-train solution (\ref{34}) also becomes the same dipole-type
soliton solution (\ref{41}). Notice that for this kind of solitons, we
obtain a relation between the pulse amplitude and duration as:%
\begin{equation}
P_{0}\tau _{0}^{2}=2\sqrt{\frac{30\epsilon }{\gamma }}.  \label{45}
\end{equation}

\noindent provided that $\gamma \epsilon >0$.

\section{Propagation of pulse-trains in extended NLSE with higher-order
dispersion}

For completeness, we have numerically simulated the propagation of
pulse-train solutions (\ref{16}), (\ref{25}) and (\ref{34}) in Eq. (\ref{1})
by utilizing the split-step Fourier method \cite{AGRA}. The physical
parameters used in this model for the case of pulse-train solution (\ref{16}%
) are representative of a femtosecond Ti: sapphire laser \cite{Piche}: $%
\beta _{2}=-100$ fs$^{2}$/cm, $\beta _{3}=-500$ fs$^{3}$/cm, $\beta
_{4}=-5000$ fs$^{4}$/cm, and $\gamma =2.36\times 10^{-6}$ W$^{-1}$cm$^{-1}$.
Note that these experimental parameters satisfy the relations $\gamma
\epsilon <0$ and $8\alpha \epsilon >3\sigma ^{2}$ which are necessary and
sufficient for the existence of the pulse-train solution (\ref{16}). As
concerns the pulse-train solutions (\ref{25}) and (\ref{34}), we use the
parameters of the silicon-based slot waveguide as \cite{Kruglov3,Roy2}: $%
\beta _{2}=-0.05$ ps$^{2}$/m, $\beta _{3}=1.9\times 10^{-5}$ ps$^{3}$/m, $%
\beta _{4}=2.5\times 10^{-5}$ ps$^{4}$/m, and $\gamma =42$ W$^{-1}$cm$^{-1}$%
. We point out that with these system parameters, the relations $\gamma
\epsilon >0$ and $8\alpha \epsilon <3\sigma ^{2}$ for the existence of
pulse-train solutions (\ref{25}) and (\ref{34}) are fulfilled. Figure 6
illustrates the evolutions of periodic pulse-trains propagating through a
distance of $20$ dispersion lengths. Here the elliptic modulus is chosen as $%
k=0.9$. We can see that the pulse-train solutions (\ref{16}), (\ref{25}) and
(\ref{34}) can preserve their shape during propagation due to the balance
between the effects of quadratic, cubic and quartic dispersions and the Kerr
nonlinearity in the system.

Finally, the figures presented above indicate that the newly found
pulse-train solutions have no oscillated tails on the wings of pulses. It
important to mention that experimental and numerical studies have
demonstrated that solitons propagating in nonlinear Kerr media with dominant
fourth-order dispersion are approximately Gaussian with oscillating tails 
\cite{Blanco,Tam}. By considering the influence of both second-, third- and
fourth-order dispersions on the pulse propagation through fibers, one of the
present authors has recently obtained quartic solitons with a $\mathrm{sech}%
^{2}$ shape which does not carry oscillating tails \cite{Kruglov2}. Under
these conditions, we have also found that dipole solitons without
oscillating tails are formed in the highly dispersive fiber system \cite%
{Kruglov3}. These important results show that dispersive effects up to the
fourth order and Kerr nonlinearity can achieve a perfect balance under
certain conditions and lead to the formation of solitons without oscillating
tails. No doubt, the pulse-trains presented here may be of guidance for the
experimental generation of optical periodic waves due to their precise
nature and interesting amplitude-duration relationship.

\section{Stability of the pulse-train solutions}

To complete the discussion of the pulse-train solutions found above, we
study their stability with respect to the finite perturbations numerically.
It is crucially important to note that only solutions that are stable (or
weakly unstable) can be observed in physical experiments and used in
practical applications \cite{Choudhuri}. We mention in passing that quartic
and dipole soliton solutions to the extended NLSE (\ref{1}) are found to be
stable with respect to finite perturbations \cite{Kruglov2,Kruglov3}. Here
we perform direct numerical simulations of the model (\ref{1}) using the
split-step Fourier method \cite{AGRA}, to test the stability of pulse-train
solutions (\ref{16}), (\ref{25}) and (\ref{34}) with initial white noise, as
compared to Figs. 2(a), (b) and (c), respectively. As usual, we put the
noise onto the initial profile, then the perturbed pulses read \cite{He}: $%
\psi _{\text{pert}}=\psi (\tau ,0)[1+0.1\,$ random$(\tau )]$. Figures 7(a),
(b) and (c) depict the numerical evolution of nonlinear wave solutions under
the perturbation of white noise. From these figures, we can see that the
pulse-train waveforms can propagate in a stable way under the initial
perturbation of white noise. Thus, we can conclude that the solutions we
obtained are stable. 
\begin{figure}[h]
\includegraphics[width=1.3\textwidth]{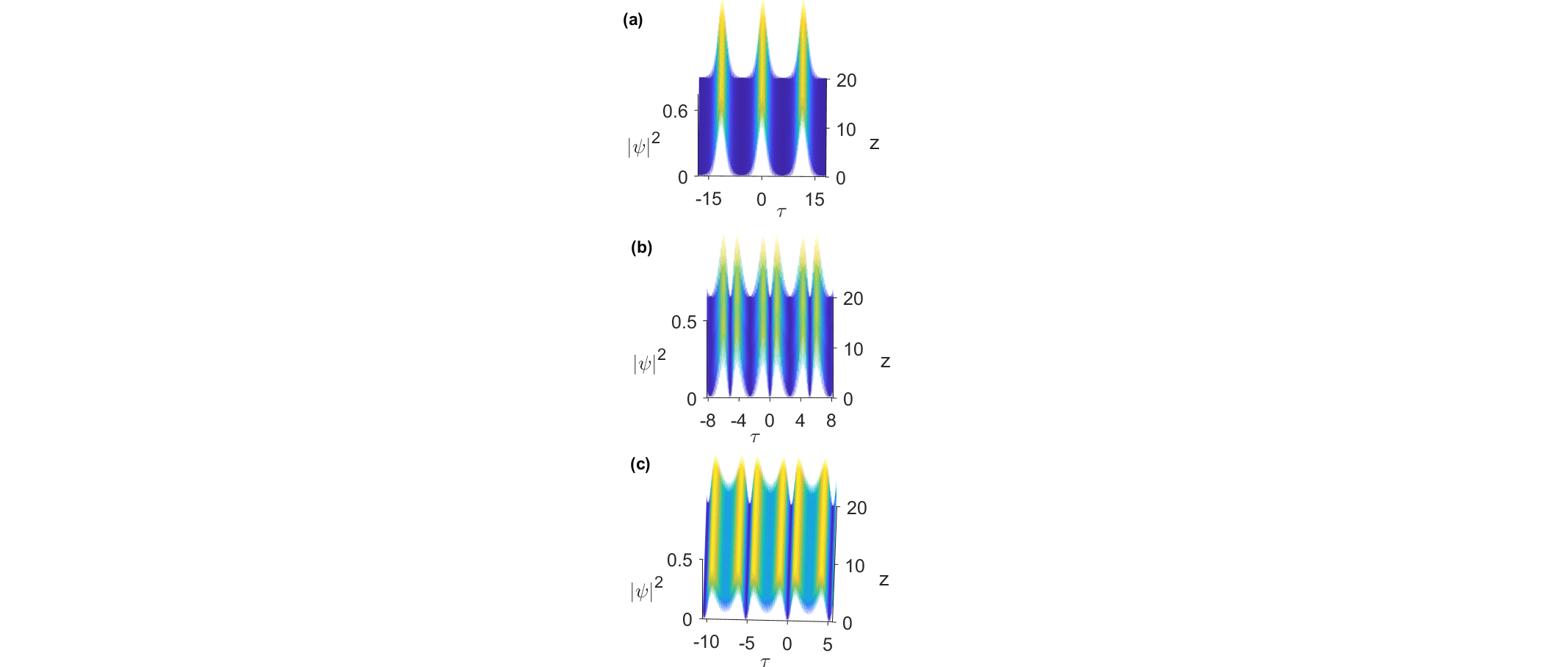}
\caption{The numerical evolution; (a) pulse-train solution (\protect\ref{16}%
), (b) pulse-train solution (\protect\ref{25}), (c) pulse-train solution (%
\protect\ref{34}) under the perturbation of white noise with maximal value
as $0.1$. The values of parameters are the same as in Figs. 2 (a), (b) and
(c) respectively.}
\label{FIG.7.}
\end{figure}

\section{Conclusion}

In conclusion, we have discovered three classes of pulse-trains whose
amplitudes, widths and wavenumbers are different with the distinctive
property of exhibiting a relation between the amplitude and duration
determined by the sign of a joint parameter uniquely. We have found that
these structures have the same velocity, which depends only on the three
dispersion parameters. It is demonstrated that higher-order dispersion may
lead to the formation of double-humped pulse-trains besides the ones taking
the usual form with single peak in each periodic unit. It is also shown that
the pulse-trains are obtained without necessarily assuming a specified
condition on the fiber parameters. Due to the precise nature, the nonlinear
waves presented here may be profitably used in designing the optimal fiber
system experiments. These results can be exploited to experimentally realize
pulse-trains in fiber systems and may help in stimulating more research in
understanding their optical transmission properties.

\end{document}